\documentclass[aps,prl,twocolumn,showpacs]{revtex4}

\usepackage{amsmath}
\usepackage{amssymb}
\usepackage{epsfig}
\usepackage{enumerate}

\newcommand{\tr}{\operatorname{tr}}

\renewcommand{\Re}{\operatorname{Re}}

\newcommand{\sub}[1]{_{\text{#1}}}

\newcommand{\bx}{\bold{x}}
\newcommand{\bq}{\bold{q}}
\newcommand{\bp}{\bold{p}}

\begin{document}

\title{Microscopic theory of the Andreev gap}

\author{Tobias~Micklitz$^1$ and Alexander Altland$^2$}

\affiliation{$^1$Materials Science Division, Argonne National Laboratory, Argonne, Illinois 60439, USA\\
$^2$Institut f\"ur Theoretische Physik, Universit\"at zu K\"oln, Z\"ulpicher Str. 77, 50937 K\"oln, Germany}
\date{\today}

\begin{abstract} 
  We present a microscopic theory of the Andreev gap, i.e. the
  phenomenon that the density of states (DoS) of normal  chaotic
  cavities attached to superconductors displays a hard gap centered around the
  Fermi energy. Our approach is based on a solution of
  the quantum Eilenberger equation in the regime $t_D\ll t_{\rm E}$, where $t_D$ and $t_\mathrm{E}$ are the
classical dwell time and Ehrenfest-time, respectively.  We
  show how quantum fluctuations eradicate the DoS at low energies and
  compute the profile of the gap to leading
order in the parameter $t_D/t_{\rm E}$. 
\end{abstract}
\maketitle

The attachment of a superconductor to a conducting cavity leads to a
suppression of the normal density of states  -- the proximity
effect. For cavities with classically chaotic dynamics, a discrepancy
is found between semiclassical calculations~\cite{prev} and such based
on random matrix theory (RMT)~\cite{rmt}: Semiclassics obtains a small
yet finite DoS for all excitation energies $\epsilon$ above the Fermi
level $\epsilon_{\rm F}$, while RMT predicts the formation of a hard
gap below some energy $\epsilon^\ast$.  The origin of this so-called
`gap problem' in Andreev billiards was pointed out by Lodder and
Nazarov some time ago~\cite{prev}: quantum corrections not captured in
the principal semiclassical approximation are expected to generate a
hard spectral gap for trajectories longer than the Ehrenfest
time. Although various semi-phenomenological realizations of this
mechanism have been formulated, a fully microscopic theory of gap
formation is outstanding. The construction of such a theory is the
goal of the present paper.

 {\it Quasiclassical Eilenberger equation} --- Consider a
two-dimensional Andreev billiard, i.e. a chaotic normal-conducting
cavity attached to a bulk superconductor. We wish to compute the
cavity DoS in a `semiclassical' regime where the
quantum time scales of the problem exceed all classical scales.  Under
these circumstances one expects~\cite{prev} the gap, $\epsilon^\ast$
to be set by the inverse of the Ehrenfest time, $\epsilon^\ast =
\pi \hbar/2 t_{\rm E}$, where $t_{\rm E} = \lambda^{-1}
\ln(c^2/\hbar)$, $\lambda$ is the dominant Lyapunov exponent of the
system, and $c^2$ a classical action scale whose detailed value is of
little relevance. Heuristically, $t_{\rm E}$ is the time a minimal
wave package needs to spread over classical portions of phase space;
the dynamics at time scales beyond $t_{\rm E}$ is no longer classical.

To compute the DoS, we start out from the quantum Eilenberger equation
(for notational convenience we suppress the infinitesimal imaginary
increment in $\epsilon+i0$) 
\begin{align}
\label{eq:1}
 \left[ \epsilon \sigma_3 - i\Delta\sigma_2+H\openone\stackrel{\ast}{,}G\right] =0
\end{align}
for the quasiclassical retarded matrix Green function, $G(\bx)$, i.e.
the Wigner transform of the Gorkov superconductor Green function.  In
(\ref{eq:1}), $\sigma_i$ are Pauli matrices acting in particle-hole
space, $\bx=(\bq,\bp)^T$ is a phase space point in the shell of
constant energy, $H(\bx)=\epsilon_F$, $H(\bx)$ is the Hamilton
function, and the order parameter amplitude $\Delta=\Delta(\bq)$ is
non-vanishing only at the cavity-superconductor interface.  The Green
function is subject to the nonlinear constraint $G\ast G=\openone$,
and yields the DoS as $\nu(\epsilon)= \frac{\nu_0}{2\Omega} \Re \int
d^2x \tr \left[ G(\bold{x})\sigma_3\right]$, where
$\Omega=\int_{H(\bx)=\epsilon_F} d^2x \,1$ is the volume of the energy
shell and $\nu_0$ the normal metallic DoS.  Finally, the symbol
`$\ast$' indicates that all products between phase space functions in
Eq.~(\ref{eq:1}) are Moyal products 
 $(A\ast B)(\bold{x}) = 
\exp\big({i\hbar\over 2} \partial_{\bold{x'}}^T
  I \partial_{\bold{x}}\big)\big|_{\bold{x}=\bold{x'}} A(\bold{x'})  B(\bold{x})$.

\begin{figure}[htbp]
   \centering
   \resizebox{3.5in}{!}{\includegraphics{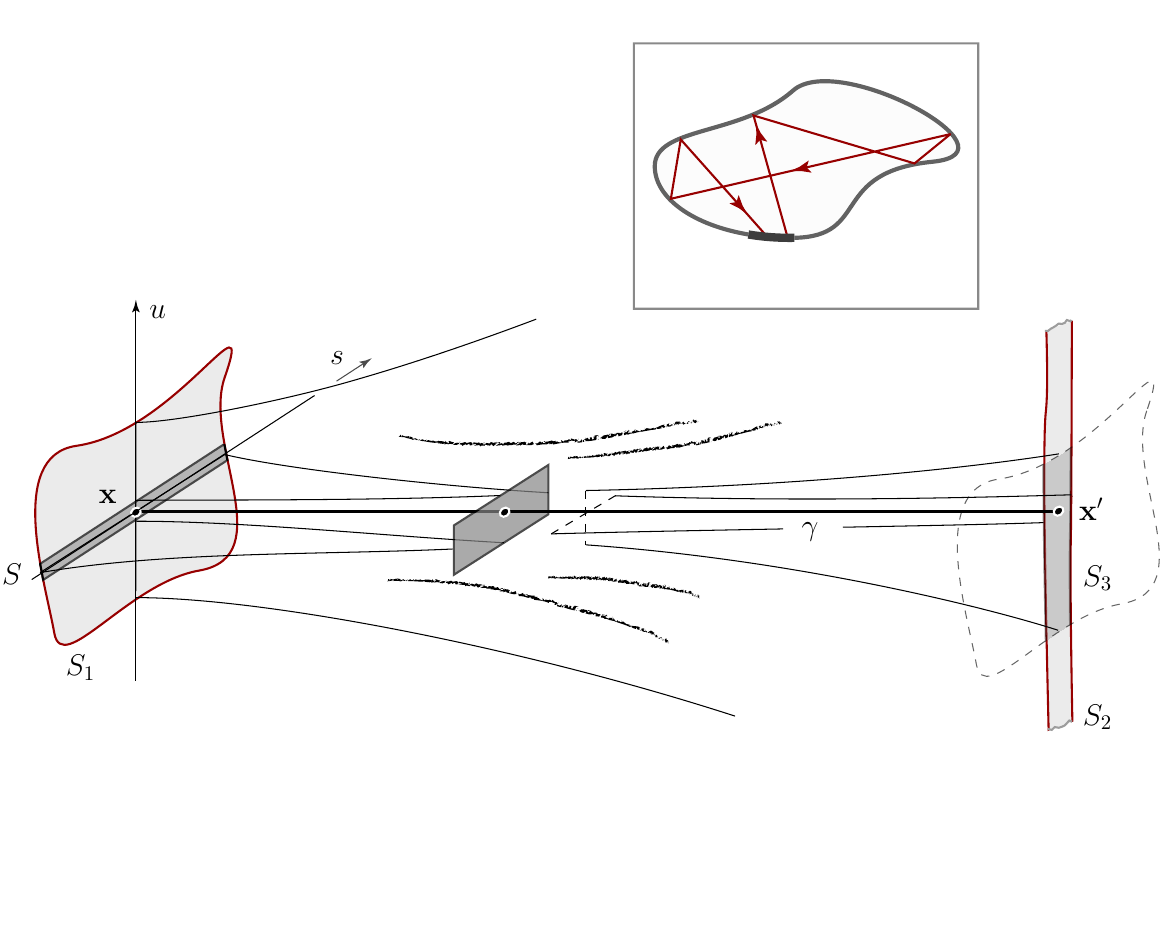}}
  \vspace{-2cm} 
  \caption{ Inset: classical trajectory connecting the
    superconductor/normal conductor interface of a chaotic Andreev
    billiard with itself. Main part: abstract phase space
    representation of that trajectory and its vicinity in a system of
    locally stable ($s$) and unstable ($u$) coordinates. The meaning
    of the shaded areas is explained in the main text.}
   \label{andreev}
 \end{figure} 
 
 {\it Classical evolution and its inconsistency}--- Upon Taylor
 expansion to lowest orders $(A\ast B)(\bx) = A(\bx) B(\bx) + {i\hbar
   \over 2}\{A,B\}(\bx) + {\cal O}(\hbar^2\partial_x^2)$ ($\{\,, \}$
 is the Poisson bracket) Eq.~(\ref{eq:1}) assumes the standard form of
 the classical Eilenberger equation~\cite{eilenberger} 
\begin{equation}
  \label{eq:5}
  [ i\epsilon \sigma_3+ \Delta\sigma_2,G]- \hbar {\cal L}G=0,  
\end{equation} 
where ${\cal L}=\{H,.\}$ generates the classical Liouville flow.  However,
(finite order) Taylor expansions of the Moyal product become
problematic in cases where the function $G$ displays structure on {\it
  linear} scales $\lesssim {\cal O}(\hbar)$ and higher order derivatives 
  ${\cal O}(\hbar^2\partial_x^2)$ become of the same order as $\hbar {\cal L}$; 
  as we shall see, this is precisely what happens on the solutions $G$ supporting the 
DoS in the region of the spectral gap.

The classical Eilenberger equation (\ref{eq:5}) describes the
evolution of $G$ along individual classical trajectories $\gamma$
beginning and ending at the superconductor interface (cf. inset of
Fig.~\ref{andreev}.) Parameterizing a trajectory $\gamma$ of length
$T$ in terms of a coordinate $t\in [-T/2,T/2]$, the Liouville
operator on $\gamma$ assumes the form ${\cal L}=\partial_t$ and
the solution in the asymptotic limit $\epsilon/\Delta\rightarrow 0$
is~\cite{prev}
 \begin{align}
 \label{eq:6}
 G_T(t)=-i\tan\left({\epsilon T\over \hbar}\right) \sigma_3 +  {
   \cos\left({2\epsilon t\over\hbar }\right) \sigma_2 +
   \sin\left({2\epsilon t\over\hbar }\right) \sigma_1 \over
   \cos\left({\epsilon T\over \hbar}\right)}.
\end{align} 
Denoting the
$\sigma_i$-components of $G$ by $G_i$, the solution obeys the boundary
conditions~\cite{prev}
\begin{align}
\label{bc}
G_{T,1}(\pm T/2)=\pm i G_{T,3}(\pm T/2), \quad G_{T,2}(\pm T/2)=1.
\end{align} 
The 
component $G_{T,3}=\mathrm{const.}$ generates
(via the identity $\text{Im}\left(\tan\left(x+i0\right)\right)=
\pi\sum_m \delta\left(x-(m+1/2)\pi\right)$) a quantization condition, $\epsilon T = (m+{1\over2})\pi\hbar$,
$m=0,1,2,...$, for the flight times of trajectories contributing to the DoS at
energy $\epsilon$. The exponential sparsity of trajectories with $T\gg
t\sub{D}$ much larger than the average dwell time~\cite{pathdist}  then
leads to an exponential suppression of the DoS for $\epsilon\lesssim
\hbar t\sub{D}^{-1}$, but not to a gap.

In view of the continuity conditions underlying the
approximation (\ref{eq:5}), it is mandatory to explore what happens as
we transversally depart from an isolated trajectory into surrounding
phase space. To this end, it is useful to interpret each trajectory as
element of a corridor or band~\cite{trbands1,effrmt,trbands6} which is
formed by all trajectories that run through the same sequence of
scattering events.  A schematic of a band is shown in the bottom part
of Fig.~\ref{andreev}, where the straight line represents a trajectory
beginning and ending at points $\bx$ and $\bx'$ in the SN
interface. We introduce Poincar\'e sections through the trajectory,
and span them by the locally stable and unstable coordinates, $s$ and
$u$, respectively.  The shaded areas then represent the SN interface
($S_1$), the image of that area under the Hamiltonian flow after time
$t$ ($S_2$), the intersection of the image with the interface ($S_3$),
and the pre-image of the intersection ($S$), respectively.  Points in
$S$ remain compactly confined and exit at the same instance
$T$. The image of $S$ under evolution defines
a 'corridor' of sections across which the
quasiclassical solutions $G_T$ is nearly constant. While the
transverse area, $us$, of the corridor is a conserved quantity, its
shape is not.  At a given instance of time, $t$, its smallest linear
extensions is given by (cf. Fig.~\ref{andreev}) $\sim {\rm
  const.}\times {\,\rm min}(e^{-\lambda
  (T/2+t)},e^{{-\lambda(T/2-t)}})$, with a classical
proportionality constant. For trajectory times $T > t_\mathrm{E}$, that
scale may shrink below ${\cal O}(\hbar)$, and this is when
Eq.~(\ref{eq:5}) becomes problematic: at low energies, $\epsilon \sim
\hbar t_{\rm E}^{-1}$, the narrow corridors of long trajectories
$T>t_{\rm E}$ meander through the bulk of phase space, in which
trajectories are of average length $\sim t_{\rm D} \ll t_{\rm E}$ and
Green functions are `locked' to the superconductor order parameter,
$G(\epsilon) \simeq \sigma_2$. (Here and throughout, we use the
notation $\simeq$ to indicate equality up to inconsequential
corrections scaling with some positive power of $\hbar$.) The ensuing
sharp variation of the solution $G_T$ over trans-corridor
sections of quantum extension $\lesssim {\cal O}(\hbar)$ conflicts
with quasiclassical smoothness conditions required for
Eq.~(\ref{eq:5}).

Our solution to the problem proceeds in two steps: we first
transversally extend (\ref{eq:6})  to a solution of 
(\ref{eq:5}) in a  `Planck tube'~\cite{fn11}
\begin{align}
\label{eq:9}
Z=\hspace{-.4cm}\bigcup_{ -{T\over2}\leq t\leq  {T\over2}}\hspace{-.4cm} Z_t,\quad 
Z_t=\{(u,s,t)|\; |us|\leq \hbar,\;|u|,|s|<c\},
\end{align} 
centered around $\gamma$. This -- singular
-- configuration will then be the basis for the construction of a smooth
configuration $G$ that solves the \textit{quantum} equation
(\ref{eq:1}) up to corrections $\sim t_D/t_\mathrm{E} $. The
quantum $G$ displays a hard spectral gap.

\begin{figure}[b!]
   \centering
   \resizebox{3.5in}{!}{\includegraphics{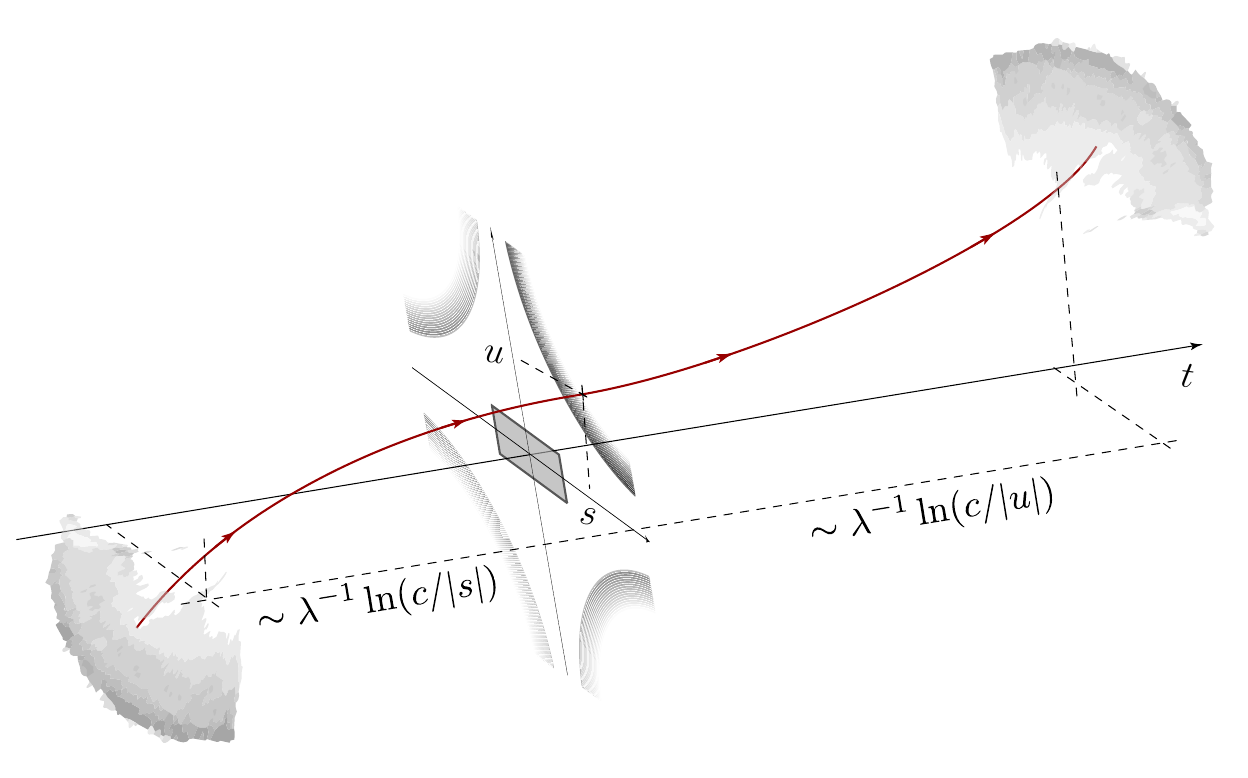}}
   \caption{Vicinity of a long trajectory in a system of locally
     stable (s) and unstable (u) coordinates. Points at the boundary 
     $|us|\sim\hbar$ belong to trajectories $\gamma$ generically of
     length $T\sim t_\mathrm{E}$ (here illustrated by the curved
     line.) The cloudy region at the ends of $\gamma$ represent
     generic phase-space points.}
\label{trj}
\end{figure} 

Consider, then, the corridor carried by a trajectory $\gamma$ of length $T\ge
t_\mathrm{E}$. (In the wide corridors of shorter trajectories the
Green function does not depend noticeably on transverse coordinates
$(u,s,t)\in Z_t$ and the solution (\ref{eq:6}) can be taken face
value.)  We assume the corridor sections $Z_t$ to be small enough to afford a
linearization~\cite{ZurekPaz}
\begin{align}
\label{eq:10}
{\cal L} =\partial_t +  \lambda u \partial_u - \lambda s\partial_s,
\end{align} 
where the terms $u\partial_u$ and $s\partial_s$ describe the
divergence and contraction of phase flow around $\gamma$,
respectively.

\begin{figure}[t!]
   \centering
   \resizebox{3.3in}{!}{\includegraphics{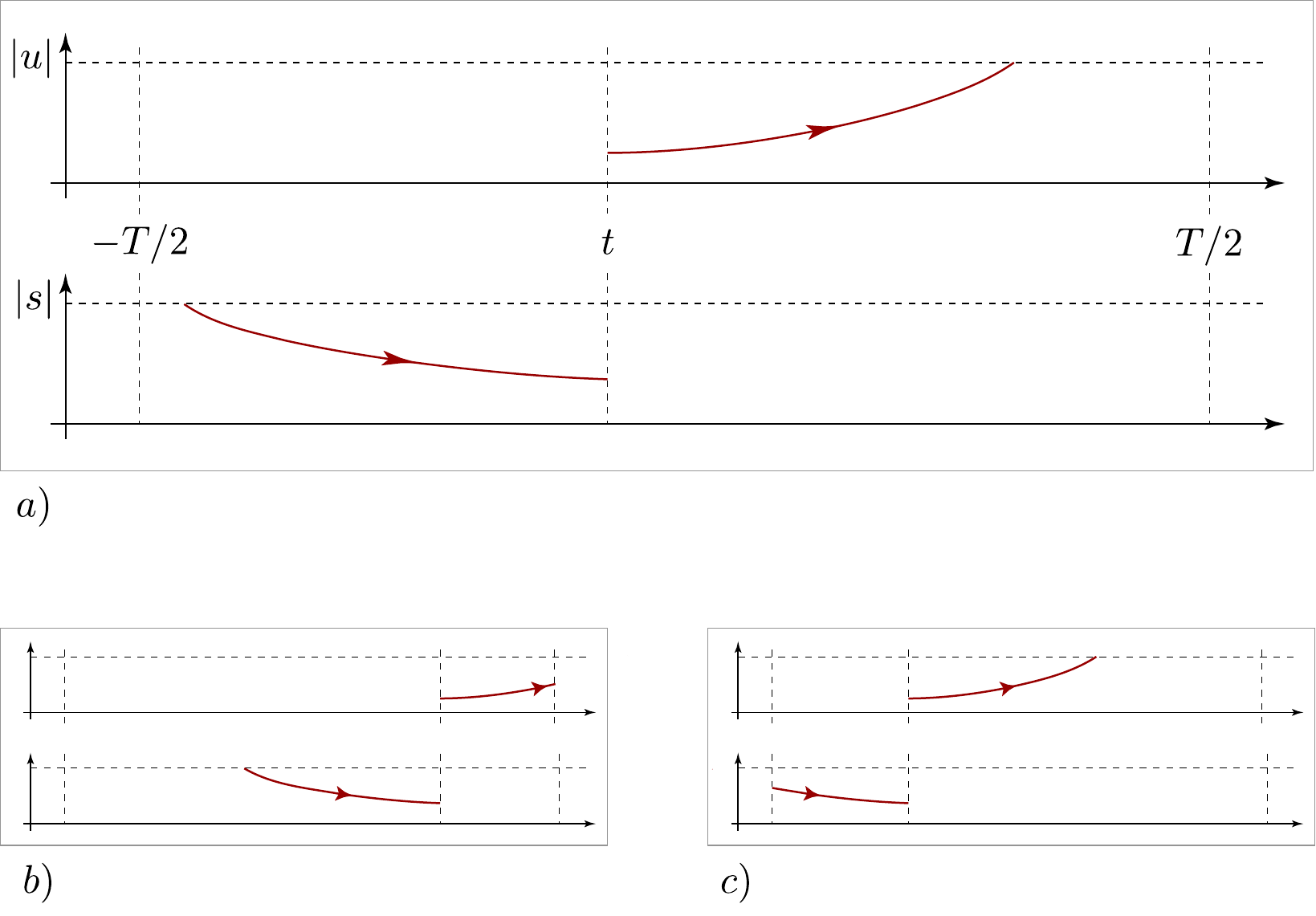}}
   \caption{
   On the length of trajectories piercing the boundary of the
   Planck cell around a long reference trajectory. a) reference times
   corresponding to a bulk point in the N-region, b) point close to
   the exit into the superconductor, c) point close to the entrance
   into the N-region. Discussion, see text.}
\label{traj_duration}
\end{figure} 
Going forward (backward) in time, the trajectory through a point
$(u,s,t)\in Z_t$ will stay in the vicinity of $\gamma$ for a time
$t(u)$ ($t(s)$) where $t(x)= \lambda^{-1}\ln(c/|x|)$.
(cf. Fig. \ref{trj}.) Thereafter a classically short time, typically
of ${\cal O}(t\sub{D})$, passes before the departing trajectory exits;
up to classical corrections, the time of flight of the trajectory
through $(u,s,t)$ thus reads $t(u)+t(s) = \lambda^{-1}\ln(c^2/|us|)$.
Specifically, for phase-space points on the boundary of the Planck
cell $|us|\sim \hbar$ and therefore $t(u)+t(s)\simeq
t_{\mathrm{E}}$. The above consideration applies to phase space points
far away from the SN interface (cf. Fig.~\ref{traj_duration} a)). For
points close to the interface, it may happen that the trajectory
through $(u,s,t)$ hits the interface before it has diverged up to $c$,
in which case the exit time is shorter than $t(u)$
(Fig.~\ref{traj_duration} b)). Or, it has been in the system for a
time shorter than $t(s)$ before the reference point is reached
(Fig.~\ref{traj_duration} c)). We subsume these different cases, by
introducing effective in- and out-times $t_o(u)=
\mathrm{min}(t(u),T/2-t)$ and $t_i(s)= \mathrm{min}(t(s),T/2+t)$,
where the function $\mathrm{min}(t,t')\equiv -{1\over \lambda}\ln
\big(e^{-\lambda t} + e^{-\lambda t'}\big)$ smoothly interpolates
between $t$ and $t'$ over a `microscopic' switching interval $\sim
\lambda^{-1}$.  These functions  evolve uniformly, in the sense ${\cal L}(t_{i/o}(s/u)) = (+/-) 1$. This means 
that the effective (up to corrections of ${\cal O}(t_D/t_\textrm{E})$)
duration of the trajectory through $(u,s,t)\in Z_t$, is given by
$T_{(u,s,t)}\equiv t_o(u) + t_i(s)$ and the trajectory parameter by
$\tau_{(u,s,t)}\equiv {1\over 2}(t_i(s) - t_o(u))$.  Substitution,
$T\to T_{(u,s,t)}$ and $t\to \tau_{(u,s,t)}$ in (\ref{eq:6}) then
obtains a transverse extension
\begin{align}
\label{eq:11}    
G^c(u,s,t)\equiv G_{T_{(u,s,t)}}(\tau_{(u,s,t)})
\end{align}
of (\ref{eq:5})~\cite{fn_bound}.  
$G^c$ solves the Eilenberger equation in direct consequence of the
flow-uniformity  of  $t_i(s)$ and $t_o(u)$. By the same token,
however, the solution becomes singular at times $T\ge t_{\mathrm{E}}$
when $t_{i,o}$ begin to display structure on scales
$\lesssim \hbar$.  Next, we show that this is not what happens 
in the full quantum dynamics.

{\it Quantum evolution and spectral gap}---  Let us define a
generalization, $t^q_{i}(u,s,t)$, of
$t_{i}(s,t)$ by requiring uniformity under the
full dynamics, $-i\hbar^{-1}[H\overset{\ast}{,}\,t_{i}^q]=1$, or
\begin{align}
\label{eq:8}
{\cal L}\,t_{i}^q
+ [{\cal
  V}\overset{\ast}{,}\,t_{i}^q]=1,
\end{align}
where $[{\cal V}\overset{\ast}{,} \;]\equiv -i\hbar^{-1}
[H\overset{\ast}{,} \;]-{\cal L}$ accounts for quantum corrections to
the linearized classical dynamics.  The above equation may be solved
by introducing `action-angle coordinates' $I=us$, $\phi\equiv {1\over
  2} \ln(u/s)$ in terms of which ${\cal L}
= \partial_t+\lambda \partial_\phi$.  The ${\cal V}$-term may now be
formally removed by `gauging' Eq.~(\ref{eq:8}) with
  \begin{align}
\label{eq:9}
U(I,\phi,t)=  {\cal P} e_\ast^{{1\over \lambda }\int_0^{\phi} d\phi' {\cal V}(I,\phi')}, 
\end{align}
where $e_\ast^{(...)}$ is defined by a Moyal series expansion in the
exponent and ${\cal
  P}$ is a $\phi$-ordering prescription (see Ref.~[\onlinecite{wl1}]
for details) accounting for the non-commutativity of ${\cal
  V}(I,\phi)$ at different values of $\phi$. By
construction~[\onlinecite{wl2}], $U$ obeys ${\cal L} U ={\cal V}\ast
U$, which means that Eq.~(\ref{eq:8}) is solved
by
\begin{align}
\label{eq:10}
t^q_i(u,s,t)=(U\ast t_i\ast U^{-1})(u,s,t).
\end{align}
 In practice,
 both the detailed form of ${\cal V}$ and of $U$ will not be
 known. This lack of knowledge, however, is not of essential concern to us; to
 the logarithmic accuracy required by the present analysis, basic
 scaling arguments suffice to determine the action of $U$ on $t_i$: 
 describing nonlinear corrections to the flow, the expansion of ${\cal
   V}$ for small $u,s$ starts as $\hbar{\cal V}=us\times {\cal O}(u^ns^m), n+m>
 0$. Accordingly, $U=1+\hbar^{-1} us\times {\cal O}(u^ns^m)$. This
 entails that 
 for any function $f$ that is smooth (analytic) around $u=s=0$,
 $(U*f*U^{-1})(u,s) = f(u,s) + {\cal O}(\partial_u f
 u^{n+1}s^m,\partial_s f u^n s^{m+1})$. At the small values of
 coordinates we are interested in, $|us|\sim \hbar$, the ${\cal
   O}(\dots)$-terms become irrelevant, which reflects the irrelevancy
 of  dynamical corrections to the linearized flow close to the
 trajectory center. 
 To explore the effect
of $U$ on singular functions (such as $t_i(s)\sim\ln(|s|)$), we notice
that for arbitrary $f(s)$, 
\begin{align}
\label{eq:14}
e^{-iku}\ast f(s) = f(s+\hbar k) \ast e^{-iku}.
\end{align} 
This identity suggests to introduce a Fourier mode decomposition 
$U(u,s,t)=\int dk \,U_{(s,t)}(k) e^{-ik u}$. Specifically, let us
consider values $|s|\sim \hbar$, where singularities begin to put the
semiclassical theory at risk.
For these values, the support of the mode coefficients
$U_{(s,t)}(k)$ extends up to 'classical' values $k\sim \hbar^0$~\cite{fn_ucl}. 
We
thus obtain $t_i^q(u,s,t) =\int dk \,t_i(s+\hbar k) \ast
F_{(s,t)}(k)$, where the positive indefinite but normalized ($\int dk
F_{(s,t)}(k)=1$) `weight' function $F_s(k) = \left(e^{-iku}
  U_{(s,t)}(k)\right)\ast U^{-1}(u,s,t)$. A straightforward estimate now shows
that for asymptotically small $\hbar$ the integral evaluates to $t_i^q(u,s,t)
= t_i(|s|+\hbar/u_0)\simeq \mathrm{min}(t_i(s),t_{\mathrm{E}})$, 
where $u_0$ is a non-universal
constant. Similarly,  $t^q_o(u,s,t)\simeq
\mathrm{min}(t_o(u),t_{\mathrm{E}}) $. 
Summarizing, we have found that the operators
$U$ act to truncate  singularities in trajectory times 
in a manner independent of the detailed form of the potential ${\cal V}$.

Building on these results it is now straightforward to construct a
smooth solution of the quantum equation (\ref{eq:1}): its general
solution is given by $C \sigma_3 + (1-C^2) \left(\cos\left({\epsilon
      \tau^q\over \hbar}\right)\sigma_2+ \sin\left({\epsilon
      \tau^q\over \hbar}\right)\sigma_1\right)$, where
$\tau^q=(t^q_i-t^q_o)/2$ and we used that smooth functions $f(\tau^q)$
('$\sin$', '$\cos$', etc.) evolve linearly, $[H\overset{\ast}{,}
\,f(\tau^q)]\simeq f'(\tau^q) [H\overset{\ast}{,}
\,\tau^q]=i\hbar f'(\tau^q)$, up to corrections of ${\cal O}(1/t_{\rm E}
\lambda)$~\cite{fn_tel}. The normalization function $C=C(u,s,t)$ is
determined by requiring stationarity $[H\overset{\ast}{,}\,C]=0$, and
compatibility with the boundary conditions~(\ref{bc}). These two
conditions lead to the identification $C=i \tan(\epsilon
T^q(u,s,t)/\hbar)$, where $T^q(u,s,t) = \min(T(u,s,t) ,
t_{\mathrm{E}})$ is the effective trajectory time. 
In conclusion, we have found that the quantum equation (\ref{eq:1}) is
solved by $G_{T^q(u,s,t)}(\tau^q(u,s,t))$, which differs from the
solution of the classical equation (\ref{eq:5}) by an upper cutoff
$t_{\mathrm{E}}$ limiting both the trajectory time $T^q$ and the
trajectory parameter $\tau^q$. Technically, this is the main result of
the present letter.

The above solution signals that quantum fluctuations couple narrow
bands of transverse extension $\lesssim \hbar$ to neighboring phase
space. This coupling is strongest in the terminal regions of long
trajectories $|\tau|\gtrsim t_{\mathrm{E}}$ where bands flatten in one
direction. Inspection of Fig.~\ref{trj} shows that the
$\hbar$-neighborhood of these segments is pierced by trajectories
whose length and parameter are uniformly given by $T\simeq
t_{\mathrm{E}}$ and $|\tau|\simeq t_{\mathrm{E}}/2$, respectively. At
these values the solutions $G$ are nearly stationary (up to
corrections ${\cal O}(t_D/t_\mathrm{E})$, and this reflects
in the asymptotic constancy of the regularized solution $G_{T^q\gtrsim
  t_{\mathrm{E}}}(\tau^q\gtrsim \pm t_{\mathrm{E}})\simeq
G_{t_{\mathrm{E}}}(\pm t_{\mathrm{E}})$ at large parameter values. The
capping of trajectory times $T^q\lesssim t_{\mathrm{E}}$ in turn
implies a vanishing of the DoS for energies $\epsilon
<\epsilon^\ast$. (Technically, $\mathrm{Re}(G_3(\bx))=- \mathrm{Im}
\tan(\epsilon T^q(u,s,t)/\hbar)$ is vanishing for these energies.) The
fact that all trajectories of nominal length $T>t_{\mathrm{E}}$
get reduced to the uniform \textit{effective} length $T^q\simeq
t_{\mathrm{E}}$ implies 
an accumulation of spectral weight at
the gap edge $\epsilon^\ast = \pi \hbar/2 t_{\mathrm{E}}$. A
straightforward estimate based on the classical density of long
trajectories $p(T) dT \sim \exp(-T/t_{\mathrm{D}}) dT$ shows that the
peak is of moderate height
$\rho(\epsilon^\ast)/\rho_{\mathrm{cl}}(\epsilon^\ast)={\cal
  O}(1)$, where $\rho_{\mathrm{cl}}(\epsilon^\ast)$ is the
semiclassical estimate of the DoS. Its width is of ${\cal
  O}(t_{\mathrm{D}}/t_{\mathrm{E}}^2)$ which reflects an uncertainty
in the effective trajectory times of ${\cal O}(t_{\mathrm{D}})$.
In a real environment, the position of the gap may also 
be affected by mesoscopic fluctuations of system parameters~\cite{silvestrov}.
However, such effects are beyond the scope of the present paper.

{\it Summary and discussion} ---
We have solved the quantum Eilenberger equation to 
leading order in the small parameter $t_D/t_E$. Our solution 
verifies the existence of a gap in the DoS of clean 
chaotic Andreev billiards. 
It is worthwhile to compare these results to two earlier approaches to 
dealing with the  singularities of the classical Eilenberger theory: 
in~[\onlinecite{effrmt}], Silvestrov {\it et al.} argued that on
bands narrower than a Planck cell, classical dynamics may be
effectively replaced by RMT modeling. 
In~[\onlinecite{vavlarkin}]  Vavilov and
Larkin, coupled the system to artificial short range disorder, fine
tuned in strength to  mimic
quantum corrections to classical propagation. 
This latter procedure renders the long time dynamics  effectively
stochastic, thus preventing the build-up of sharply defined phase space structures.
Our analysis shows that phenomenological input of either type is not, in fact,
necessary. The conjunction of classical hyperbolicity and quantum
uncertainty encoded in the native Eilenberger equation automatically
regularizes classical singularities at large times. This mechanism
operates under rather general conditions and can be described at
moderate theoretical efforts. We therefore believe  the
concepts discussed above to be of wider applicability.

We are grateful for discussions with P.~Brouwer. This work was supported 
by SFB/TR 12 of the Deutsche Forschungsgemeinschaft and the
U.S. Department of Energy, Office of Science, under Contract
No. DE-AC02-06CH11357.

\pagestyle{empty}


\begin{thebibliography}{}
\bibitem{pathdist}
The distribution of path lengths $T$ in chaotic billiards is
$p(T)\propto e^{-T/t_{\rm D}}$~\cite{dt}.
\bibitem{fn_ucl} The (super)linear scaling of ${\cal V}$ in $s$
implies that as a function of $u$, ${\cal V}(u,s\sim \hbar)\lesssim
\hbar^0$. By the same token, the $u$-dependence of the function
$U(u,s\sim\hbar)\sim \hbar^0$ is classical, and the same applies to
the mode spectrum.
\bibitem{fn_tel} This is seen by straightforward comparison of the
  action of the Moyal product's higher
  order derivatives  on the envelope
  function $f$ and its argument function $\tau^q$, resp.
\end{thebibliography}
\end{document}